\documentstyle[pra,aps,twocolumn,epsfig]{revtex}
\begin{document}
\hyphenation{Ryd-berg}
\flushbottom
\draft
\title{Excitation of weakly bound Rydberg electrons by half-cycle pulses}
\author{O.~Zobay$^1$ and G.~Alber$^2$} 
\address{$^1$Optical Sciences Center, University of Arizona,
Tucson, Arizona 85721\\
$^2$Abteilung f\"ur Quantenphysik, Universit\"at Ulm,
89069 Ulm, Germany\\
to be published in {\bf Phy. Rev. A}}
\maketitle
\begin{abstract}
The interaction of a weakly bound Rydberg electron with an electromagnetic
half-cycle pulse (HCP) is described with the help of a multidimensional
semiclassical treatment. This approach
relates the quantum evolution of the electron to its
underlying classical dynamics. The method is nonperturbative and is valid
for arbitrary spatial and temporal shapes of the applied HCP.
On the basis of this approach angle- and energy-resolved spectra
resulting from the ionization of Rydberg atoms by HCPs are analyzed.
The different types of spectra obtainable in the sudden-impact approximation
are characterized in terms of the appearing semiclassical scattering
phenomena. Typical modifications of the spectra originating from finite
pulse effects are discussed.
\end{abstract}
\narrowtext
 
\section{Introduction}
Recent experimental \cite{JYB93,J96,RCSB96} and theoretical
\cite{RSB94,MC96,Delospaper} investigations have demonstrated
that (almost) unipolar, high power electromagnetic field pulses are 
a useful new spectroscopic tool which is particularly well suited 
for investigating the dynamics of weakly bound Rydberg electrons. In current
experiments the duration $\tau$ of these half-cycle pulses (HCPs) extends
from the subpicosecond to the nanosecond regime.  
Several features distinguish the interaction of Rydberg electrons with this
new type of electromagnetic radiation from the interaction with conventional
laser light or microwave fields.
First of all, and in contrast to the optical case, the HCP
interacts with the Rydberg electron at every point of its orbit.
Furthermore, due to its unipolar nature the influence of the
HCP on the electron can often be described in relatively simple and
intuitive terms,
e.g., with the help of the so-called sudden-impact approximation where
the effect of the HCP is modeled as an instantaneous momentum change.
Finally, the energy transfer between an
oscillating electromagnetic field and a Rydberg electron is always governed
by an approximate energy conservation whose energy uncertainty is typically
small in comparison with the amount of energy transferred. As HCPs are
almost unipolar such an approximate energy conservation does not hold and
arbitrary amounts of energies can be transferred to a charged particle from
zero up to a maximum amount of the order of $\hbar/\tau$.

These characteristic features give rise to interesting novel phenomena
which have been explored in a number of recent investigations.
So far these studies have concentrated
on total and energy-resolved ionization probabilities
\cite{JYB93,J96,RCSB96,RSB94,MC96,Delospaper}.
Theoretical work in this context has focused on numerical approaches
\cite{RSB94} that often employ the sudden-impact
approximation and on one-dimensional models \cite{MC96,Delospaper}
which try to capture the most significant physical effects
occuring along the direction of
polarization of a linearly polarized HCP.

In this paper a multi-dimensional semiclassical description
of excitation of a weakly bound Rydberg electron by a
half-cycle pulse is developed. It quantitatively connects the quantum evolution
of the Rydberg electron to its underlying classical dynamics.
The presented theoretical treatment is nonperturbative and it is applicable
to arbitrary spatial and temporal HCP shapes.
In particular, the approach is not restricted to
the sudden-impact approximation. Within this theoretical framework
measurable transition probabilities are represented as coherent sums of
probability amplitudes which are associated with corresponding classical
trajectories of the excited Rydberg electron.
This method is very accurate numerically especially for
highly excited Rydberg states with large principal quantum numbers and for
HCPs whose transferred energy is small in comparison with typical
ground-state ionization energies.
It is precisely this dynamical regime which usually causes severe numerical
problems in fully quantum mechanical calculations due to the large
spatial extension of Rydberg states and the presence of the
Coulomb singularity close to the nucleus.

On the basis of this theoretical treatment we demonstrate
that angular distributions of ionization probabilities contain a wealth of
information about the ionization dynamics of the Rydberg electron 
which cannot be obtained from energy-resolved ionization spectra alone.
In the sudden-impact approximation, for example,
in which the pulse duration of the exciting HCP is small
in comparison with the classical Kepler period of the Rydberg electron
energy- and angle-resolved ionization spectra are dominated by oscillatory
structures and semiclassical catastrophes of the rainbow and glory type.
These structures can be explained in a clear and intuitive way in terms of
interferences between probability amplitudes of a few classical electronic
trajectories. The study of the influence of finite pulse durations 
indicates that some parts of these oscillatory structures depend
strongly on the duration of the exciting HCP.
A main purpose of our subsequent discussion is thus the systematic exploration
of these latter effects which cannot be described appropriately by
the sudden-impact approximation. First results of our studies have
also been presented in Ref.\ \cite{AZ98}.

This paper is organized as follows:
In Sec.\ II a general (multidimensional) semiclassical description of
excitation of a weakly bound Rydberg electron by an HCP is developed.
This general approach is valid for arbitrary spatial and temporal dependences
of the HCP. Starting from this general approach the simplifications arising
in the sudden-impact approximation are discussed in detail.
In Sec.\ III we first present a characterization of the various types
of energy- and angle-resolved ionization spectra that can be obtained in
the sudden-impact approximation. These spectra may be classified
according to the appearing semiclassical scattering phenomena. 
Subsequently, it is investigated how a finite duration of the exciting HCP
influences the behavior of the spectra.
The comparison with the corresponding results obtained in the sudden-impact
approximation demonstrate which dynamical aspects are particularly sensitive
to pulse duration effects. Finally, in Sec.\ IV a brief summary and
conclusions are given.

\section{Theoretical description}

In the first part of this section a nonperturbative semiclassical framework
is developed that describes the excitation of a weakly bound Rydberg electron
by an HCP and that is valid for arbitrary spatial and temporal pulse
shapes. In the second part this treatment is specialized to the case of
short HCPs whose influence on a Rydberg electron can be described within
the sudden-impact approximation. Hartree atomic units
are used with $\hbar=m_{\rm e}=e=1$ ($m_{\rm e}$ and $e$ denote the electronic
mass and charge).

\subsection{General multidimensional semiclassical description}

Let us consider a typical process in which a weakly bound Rydberg electron
is excited by a half-cycle pulse.
The dynamics of the electronic wavefunction $|\psi\rangle_t$ is described
by the
time-dependent Schr\"odinger equation $id|\psi\rangle_t/dt = H |\psi\rangle_t$
with the Hamiltonian
\begin{equation}
H = \textstyle{\frac{1}{2}}[-i\nabla_{{\bf{x}}} - {\bf{A}}({\bf{x}},t)]^2 
+ V({\bf{x}}).
\label{Schr}
\end{equation}
Outside the core region the effective potential $V({\bf{x}})$
experienced by the Rydberg electron is of the form
$V({\bf{x}})\simeq -1/|{\bf x}|$ whereas inside
it is modified due to the presence of the other electrons and of the nucleus.
The electromagnetic vector potential 
of the exciting HCP is denoted ${\bf{A}}({\bf{x}},t)$ and fulfills the
relations $\nabla \cdot{\bf{A}}({\bf{x}},t) = 0$ and
${\bf{A}}({\bf{x}},t \to -\infty) = 0$.

Initially the Rydberg electron is supposed
to be prepared in an energy eigenstate
$|n_0 l_0 m_0\rangle$ with principal quantum number
$n_0 \gg 1$ and angular quantum numbers $l_0$ and $m_0$.
Well inside the classically allowed region and for low
values of the angular momenta, i.e. $l_0, m_0 \ll n_0$, this state
can be approximated semiclassically by \cite{BS}
\begin{eqnarray}
\langle {\bf x} | n_0 l_0 m_0 \rangle &=& A_{cl}({\bf x})[
e^{iS_0 ({\bf{x}}) - i\pi/4} + 
e^{-iS_0 ({\bf{x}}) + i\pi/4}]
\label{initial}
\end{eqnarray}
with the classical eikonal
\begin{equation}
S_0 ({\bf{x}}) = \int_0^{|{\bf{x}}|}dr' p(r',\epsilon_0) -
(l_0 + 1/2)\pi + \pi \alpha.
\label{eikonal}
\end{equation}
The local radial electronic momentum in the Coulomb potential of the ionic core
is given by
\begin{equation}
p(r',\epsilon_0) = \sqrt{2(\epsilon_0 + 1/r')}
\label{momentum}
\end{equation}
with the initial energy
\begin{equation}
\epsilon_0 = -[2(n_0 - \alpha)^2]^{-1}.
\end{equation}
While the rapidly oscillating part of the wave function (\ref{initial})
is determined by the eikonal $S_0 ({\bf{x}})$ its slowly varying amplitude
is given by
\begin{equation}
A_{cl}({\bf{x}}) = \frac{Y_{l_0}^{m_0}(\Theta,\Phi)}
{|{\bf{x}}|(n_0 - \alpha)^{3/2}\sqrt{2\pi p(|{\bf{x}}|,\epsilon_0)}}
\end{equation}
with the spherical harmonic $Y_{l_0}^{m_0}(\Theta,\Phi)$. The spherical
angles of ${\bf{x}}$ are denoted $\Theta$ and $\Phi$, respectively.
Equation (\ref{initial}) is valid for electronic distances $r=|\bf{x}|$
from the nucleus which are located well outside the core region. All
electron correlation effects are localized inside this core region which
has an extension of a few Bohr radii. They can be taken into account 
within the framework of quantum defect theory \cite{Seaton,Fano}. In the
case of an inert ionic core which is considered here for the sake of
simplicity these core effects can be described by a single
quantum defect $\alpha$. This quantum defect
is approximately energy independent sufficiently close to the ionization
threshold. Generalizations to more complicated situations in which electronic
core excitations and channel couplings have to be taken into account are
possible within the framework of multi-channel quantum defect theory
\cite{Seaton,Fano} but will not be considered in this work.

In Eq.~(\ref{initial}) the initial state $|\psi\rangle_{t=0}$
is represented by a sum of contributions each of which involves a
slowly varying amplitude, i.e.\ $A_{cl}({\bf{x}})$,
and an exponential function with a large,
imaginary-valued argument, i.e.\ $iS_0({\bf{x}})$.
Within the semiclassical treatment one associates with each of these
contributions a three-dimensional Lagrangian manifold 
in phase space \cite{Maslov,Delos}.
The special form of the initial condition of
Eq.~(\ref{initial}) implies that this Lagrangian manifold has two branches,
namely
\begin{equation}
L_0^{\pm}=
\{({\bf{x}}_0,{\bf{p}}_0)\in {\bf R}^6\mid {\bf{p}}_0 =
\pm\nabla_{{\bf{x}}_0} S_0({\bf{x}}_0) \in {\bf R}^3\}.
\end{equation}
As $S_0({\bf{x}}_0)$ depends only on the radial electronic distance
$r_0 =|{\bf{x}}_0|$ one finds ${\bf{p}}_0 = \pm |{\bf{p}}_0|
{\bf{x}}_0/| {\bf{x}}_0|$ as long as $n_0 \gg l_0, m_0$.
Semiclassically, the solution of the time-dependent Schr\"odinger equation
$|\psi\rangle_{t}$
is determined by all solutions 
${\bf{x}}_{\pm}(t;{\bf{x}}_0,{\bf{p}}_0)$ and
${\bf{p}}_{\pm}(t,{\bf{x}}_0,{\bf{p}}_0)$
of the classical equations of motion with Hamiltonian 
\begin{eqnarray}
H_{cl} &=& \textstyle{\frac{1}{2}}[{{\bf{p}}} - {\bf{A}}({\bf{x}},t)]^2 
+ V({\bf{x}})
\label{Hamclass}
\end{eqnarray}
whose initial conditions are located in this Lagrangian manifold $L_0^{\pm}$.
The probability amplitude $\langle {\bf{x}}|\psi\rangle_t$ is
determined by all those classical trajectories
$\{ {\bf{x}}^{(j)}_{\pm}(t;{\bf{x}}^{(j)}_0,{\bf{p}}^{(j)}_0)$,
${\bf{p}}^{(j)}_{\pm}(t,{\bf{x}}^{(j)}_0,{\bf{p}}^{(j)}_0)\}$
which reach point
${\bf{x}}$ at time $t$, i.e. \cite{Maslov}
\begin{eqnarray}
\langle {\bf{x}}|\psi\rangle_t &=&\sum_j 
\left(\left|
\frac{dx\wedge dy \wedge dz}
{dx_0\wedge dy_0 \wedge dz_0}\right|_t^{-1/2}
\right)_j\times\nonumber\\
&&
e^{i[S_j({\bf{x}},t) - \pi\mu_j/2]}\langle{\bf{x}}_0^{(j)}|\psi\rangle_{t=0}.
\label{solution}
\end{eqnarray}
For each contributing trajectory $j$ this probability amplitude is determined
by three characteristic classical properties, namely
its classical action 
\begin{eqnarray}
S_j({\bf{x}},t) &=&\int_0^t dt'{\cal L}(\dot{\bf{x}}^{(j)}(t'),{\bf{x}}^{(j)}
(t'),t'),
\end{eqnarray}
the determinant of its Jacobi field \cite{Schulman}
\begin{equation}
\left(\left. \frac{dx\wedge dy \wedge dz}
{dx_0\wedge dy_0 \wedge dz_0}\right|_t \right)_j
\equiv
\left(\left. {\rm Det}\frac{\partial (x,y,z)}
{\partial (x_0,y_0,z_0)}\right|_t \right)_j,
\end{equation}
and its Morse index $\mu_j$.
Thereby, ${\cal L}(\dot{\bf{x}},{\bf{x}},t)$ denotes the classical Lagrange
function associated with the Hamiltonian of Eq.~(\ref{Hamclass}).
The determinant of the Jacobi field characterizes
the stability properties of the contributing classical trajectory.
The Morse index $\mu$ is numerically equal to the numbers of zeroes of this
determinant times their multiplicities \cite{Schulman}.

From Eq.~(\ref{solution}) one can evaluate the probability amplitude
of measuring an ionized electron with final momentum ${\bf{p}}^{(f)}$.
With the help of the stationary phase approximation one obtains the result
\begin{eqnarray}
\langle {\bf{p}}^{(f)}|\psi\rangle_{t\to\infty} &\equiv&
\lim_{t\to\infty}(2\pi)^{-3/2}\int d^3{\bf{x}}\,
e^{i[\epsilon^{(f)}t -
{\bf{p}}^{(f)}\cdot{\bf{x}}]}\langle {\bf{x}}|\psi\rangle_t 
\nonumber\\
&=&
\sum_j P_j^{(cl)} 
e^{i[S_j({\bf{p}}^{(f)}) + W_j({\bf{x}}^{(j)}_0) - i\pi{\tilde \mu}_j/2]}.
\label{semi}
\end{eqnarray}
This ionization amplitude is represented as a sum of all trajectories
$j$ which start at positions ${\bf{x}}^{(j)}_0$ with momenta
${\bf{p}}^{(j)}_0$ and which assume the final momentum ${\bf{p}}^{(f)}$ 
long after the interaction with the HCP, i.e., at $t \to \infty$.
In Eq.~(\ref{semi}) the quantity
\begin{eqnarray}
P_j^{(cl)} &=& 
A_{cl}({\bf x}^{(j)}_0)\left| \left(
\frac{dp_x^{(f)} \wedge dp_y^{(f)}\wedge dp_z^{(f)}}
{dx_0 \wedge dy_0 \wedge dz_0}\right)_j^{-1/2} \right|
\end{eqnarray}
denotes the classical contribution of trajectory $j$
to the angle- and energy-resolved
ionization probability.
The classical action $S_j({\bf{p}})$ is given by
\begin{eqnarray}
S_j({\bf{p}}) &=& - \int_0^{\infty}\! dt\, {\bf{x}}^{(j)}(t)\cdot
\frac{d{\bf{p}}^{(j)}(t)}{dt}
- {\bf{x}}^{(j)}(t=0)\cdot{\bf{p}}^{(j)}(t=0).\nonumber
\end{eqnarray}
The phase contribution originating from the initial state is determined by
\begin{eqnarray}
W_j({\bf{x}}) &=& \pm [S_0({\bf{x}}) - \pi/4] 
\end{eqnarray}
where one has to choose the plus or minus sign depending on whether the
initial radial momentum of the Rydberg electron is positive or negative. The
Maslov index 
${\tilde \mu}_j$ is equal to the number of zeroes of the determinant
$\left(\frac{dp_x^{(f)} \wedge dp_y^{(f)}\wedge dp_z^{(f)}}
{dx_0 \wedge dy_0 \wedge dz_0}\right)_j$ times their multiplicities
\cite{Maslov}. In terms of the ionization amplitude of Eq.~(\ref{semi})
the energy- and angle-resolved ionization probability is given by
\begin{eqnarray}
\frac{d P_{ion}}{d\epsilon^{(f)}  d\Omega} &=& \sqrt{2\epsilon^{(f)}}\mid
\langle {\bf{p}}^{(f)}|\psi\rangle_{t\to \infty}
\mid^2
\label{diff}
\end{eqnarray}
with $d\Omega = \sin\Theta_f d\Theta_f d\Phi_f$ and with
$\Theta_f$ and $\Phi_f$ denoting the spherical angles of the final momentum
${\bf{p}}^{(f)}$.

Equations (\ref{semi}) and (\ref{diff})
are main results of this section. They yield a complete and
numerically accurate (compare with Sec.\ III A) semiclassical
description of the influence of an arbitrary HCP on a weakly bound
Rydberg electron that is prepared initially in an energy eigenstate.
These equations are based on
Maslov's multidimensional
generalization of the JWKB method as applied
to the time-dependent Hamiltonian of Eq.\ (\ref{Schr}) \cite{Maslov}.
It should be mentioned that there exists also a wealth of alternative
semiclassical methods to approach initial value problems. These methods
have been pioneered by E.\ Heller \cite{H1} and W.~H.\ Miller \cite{M1}
and have been generalized recently in various directions \cite{HK,K,CB}.
These types of semiclassical approximations are mainly based
on the dynamics of Gaussian wave packets \cite{Grossmann} and
so far they have been applied predominantly
to problems with explicitely time-independent Hamiltonians.

If the initial state of the Rydberg electron is not an energy
eigenstate of the form of Eq.~(\ref{initial}) almost all
of the considerations of this section still apply. It is only the classical
Lagrangian manifold associated with the initial state that has
to be modified appropriately. In the case of a spatially localized
electronic Rydberg wave packet, for example, which can be described by
a quantum state of the form
\begin{eqnarray}
\langle {\bf{x}}|\psi\rangle_{t=0} &=&
B({\bf{x}}) e^{iR({\bf{x}})}
\label{wavepacket}
\end{eqnarray}
with a large eikonal $R({\bf{x}})$ and a slowly varying amplitude
$B({\bf{x}})$ the associated Lagrangian manifold is given by
\begin{eqnarray}
L_0 &=&
\{({\bf{x}}_0,{\bf{p}}_0)\in {\bf R}^6| {\bf{p}}_0 =
\nabla_{{\bf{x}}_0} R({\bf{x}}_0) \in {\bf R}^3\}.
\end{eqnarray}
Contrary to the initial state considered in Eq.~(\ref{initial})
this Lagrangian manifold has only one branch so that the structure of the
interferences appearing in Eq.~(\ref{semi}) is expected to
be changed significantly.

Finally, we would like to mention a useful
scaling property of the classical Hamiltonian
(\ref{Hamclass}). If spatial variations of the pulses are
negligible the classical trajectories $({\bf x}(t),{\bf p}(t))$ and
$(\tilde{\bf x}(t),\tilde{\bf p}(t))$ induced by two
HCPs ${\bf A}(t)$ and $\tilde{\bf A}(t)=\gamma {\bf A}(\gamma^3t)$ 
are related by the scaling relation
\begin{eqnarray}
\tilde{\bf x}(t) &=&\gamma^{-2}{\bf x}(\gamma^3 t),\nonumber\\
\tilde{\bf p}(t) &=&\gamma{\bf p}(\gamma^3 t).
\label{scaling}
\end{eqnarray}
These scaling properties are exploited in Sec.\ III A where they
simplify the classification of the types of spectra obtainable in the
sudden-ionization approximation.

\subsection{Semiclassical treatment in the sudden-impact approximation}

If the pulse duration $\tau$ of the exciting HCP is small in comparison with
the classical Kepler period $T_{cl} = 2\pi (n_0 - \alpha)^3$ of the initially
prepared Rydberg electron, the determination of $|\psi\rangle_t$ can be
simplified considerably. Neglecting for the sake of clarity any spatial
dependence the vector potential of the applied HCP can be approximated by
\begin{eqnarray}
{\bf A}(t) &=& - \Theta(t) \Delta{\bf p}
\end{eqnarray}
with the characteristic momentum
\begin{eqnarray}
\Delta {\bf p} &=& - \int_{-\infty}^{\infty} dt\, {\bf{E}}(t).
\end{eqnarray}
Thereby, ${\bf E}(t)=-d{\bf A}/dt$ denotes the HCP field strength.
Thus the solution $|\psi\rangle_t$ of the time-dependent Schr\"odinger equation
with Hamiltonian (\ref{Schr}) is approximately given by
\begin{eqnarray}
|\psi\rangle_t &=& e^{-iH_A t} e^{i\Delta{\bf p}\cdot \hat{\bf x}}
|\psi\rangle_{t=0} \hspace{1cm}(t > 0)
\end{eqnarray}
with the atomic Hamiltonian $H_A = 1/2(-i\nabla_{\bf x})^2 + V({\bf x})$.
Within this sudden-impact approximation the Rydberg electron experiences a
sudden change of its momentum by the amount $\Delta{\bf p}$ and subsequently
evolves under the influence of
the atomic Hamiltonian $H_A$ without being further affected by the HCP.
In this case ionization probabilities can also be evaluated quantum
mechanically by means of a partial wave analysis.

In the case of a hydrogen atom, for example, which is prepared initially
in the state
$|n_0,l_0=m_0=0\rangle$ the angle- and energy-resolved ionization probability
is given by the partial wave expansion
\begin{eqnarray}
&&\frac{dP_{ion}}{d\epsilon^{(f)} d\Omega} =
\frac{1}{\sqrt{2}\pi |{\bf p}^{(f)}|}
\sum_{l=0}^{\infty} e^{i\sigma_l} P_l(\cos\Theta_f)\times\nonumber\\
&&
\int_0^{\infty} dr\, S_{n_0,l_0=0}(r) j_l(|\Delta{\bf p}| r)
F_{l}(\epsilon^{(f)};r).
\label{partial}
\end{eqnarray}
Thereby $P_l$ and $j_l$ are the Legendre polynomial and the spherical Bessel
function of order $l$, respectively. The regular, energy normalized radial
Coulomb wave function
is denoted $F_l(\epsilon^{(f)};r)$ and $\sigma_l$ is the Coulomb phase shift.
The radial wave function of the initially prepared Rydberg state
$|n_0,l_0=m_0=0\rangle$ is denoted $S_{n_0, l_0=0}$. In Eq.~(\ref{partial})
it has been assumed that the HCP is linearly polarized and that the spherical
angles $(\Theta_f, \Phi_f)$
of the final momentum ${\bf{p}}^{(f)}$ are measured with respect to
the direction of polarization.

Alternatively, $|\psi\rangle_t$ can also be determined semiclassically with
the approach developed in Sec.~II A. For this purpose one starts from the
electronic state which is modified by the sudden
momentum change due to the HCP. Immediately after the application of the HCP
this state is given by
\begin{equation}
|\psi\rangle_{t=+0} = e^{i\Delta{\bf p}\cdot{\hat{\bf x}}}
|\psi\rangle_{t=-0}.
\end{equation}
If the Rydberg electron is prepared initially in the energy eigenstate
$|\psi\rangle_{t=-0}=|n_0 l_0 m_0\rangle$ of Eq.~(\ref{initial}), the
classical Lagrangian manifold which is associated with this state is given by
\begin{equation}
L_0^{\pm}=
\{({\bf{x}}_0,{\bf{p}}_0)\in {\bf R}^6| {\bf{p}}_0 =
\pm\nabla_{{\bf{x}}_0} S_0({\bf{x}}_0) + \Delta{\bf p}\in {\bf R}^3\}.
\nonumber
\end{equation}
In analogy to Eq.~(\ref{solution}) the probability amplitude
$\langle {\bf{x}}|\psi\rangle_t$ is determined by all solutions of the
classical equations of motion with Hamiltonian
\begin{equation}
H_{A,cl} =\frac{{\bf{p}}^2}{2} + V({\bf{x}})
\label{HAcl}
\end{equation}
whose initial conditions are located in this Lagrangian manifold. As this
Hamiltonian is explicitly time independent the energy of these trajectories
is conserved during their time evolution. This implies that outside the core
region where $V({\bf{x}}) \approx -1/|{\bf x}|$
the initial positions ${\bf{x}}_0$ on this Lagrangian manifold which yield
ionizing trajectories with
final energy $\epsilon^{(f)}$ are determined by energy conservation, i.e.
\begin{eqnarray}
\frac{1}{2}[\pm \nabla_{{\bf{x}}_0} S_0({\bf{x}}_0)
+ \Delta{\bf p}]^2 - \frac{1}{|{\bf x}_0 |}
&=& \epsilon^{(f)}.
\end{eqnarray}
Inserting Eq.~(\ref{eikonal}) and assuming a linearly polarized HCP with
$\Delta{\bf p}=\Delta p\,{\bf e}_z$, $\Delta p>0$, yields
\begin{eqnarray}
\frac{1}{| {\bf x}_0 |}
&=& \frac{(\epsilon^{(f)} - \epsilon_0 -
\Delta p^2/2)^2}{2\Delta p^2 \cos^2\Theta_0}-\epsilon_0
\label{energy}
\end{eqnarray}
with $\Theta_0$ denoting the polar angle of the initial position
${\bf x}_0$. Equation (\ref{energy}) determines the
initial positions ${\bf x}_0$ of the classical trajectories which ionize with
asymptotic energy $\epsilon^{(f)}$. According to Eq.~(\ref{energy}) there is
one value of $r_0 =|{\bf x}_0|$
for each final energy $\epsilon^{(f)}$ and initial angle $\Theta_0$.
At the critical final energy $\epsilon^{(f)}_{crit}=\epsilon_0 +
\Delta p^2/2$ the local radial momentum of Eq.~(\ref{momentum}) vanishes and
the primitive semiclassical approximation for the initial state
$|n_0 l_0 m_0\rangle$ is no longer applicable. In this special case a proper
semiclassical description of the initial state can be obtained
with the help of uniform semiclassical methods \cite{Berry}.
This peculiar property suggests that one can at least distinguish two
different energy regimes in the excitation of a Rydberg
electron by a HCP, namely a region of low final energies for which
$\epsilon^{(f)} < \epsilon^{(f)}_{crit}$ and a
high energy regime in which $\epsilon^{(f)} > \epsilon^{(f)}_{crit}$.
For a fixed value of the final ionization energy $\epsilon^{(f)}$ the initial
positions of the classical trajectories which ionize into a particular
final angle $\Theta_f$ have to be determined by the solutions of the classical
equations of motion with Hamiltonian (\ref{HAcl}). In the special case of
a hydrogen atom these relations can be evaluated analytically
\cite{Landau}. A comprehensive discussion of energy- and angle-resolved
ionization probabilities which is based on the semiclassical results of
this section is presented in Sec.~III.

\section{Energy- and angle-resolved ionization spectra}

In this section energy- and angle-resolved ionization probabilities are
discussed on the basis of the semiclassical approach developed in Sec.~II.
In the first part characteristic predictions of the sudden-impact
approximation are discussed in the semiclassical limit. 
The second part discusses modifications of these ionization spectra
that arise from a finite duration of the exciting HCP.

\subsection{Sudden-impact approximation}

This part gives a general overview of the characteristic properties
of angle- and energy-resolved ionization spectra in the sudden-impact
approximation for the case $l_0,m_0\ll n_0$.
As it will be shown below these spectra can be
described very accurately with the help of semiclassical methods.
Thus the characteristic qualitative aspects of their behavior may
be understood by studying the underlying classical dynamics.

The classical trajectories pertaining
to the parameter sets $(\epsilon_0,\Delta p,\epsilon^{(f)})$
and $(\gamma^2\epsilon_0,\gamma\Delta p,\gamma^2\epsilon^{(f)})$,
$\gamma>0$, are
related by the scaling transformation of Eq.\ (\ref{scaling}).
Therefore it is sufficient to examine
the classical dynamics as a function of the parameters $\Delta
p^2/2|\epsilon_0|$ and $\epsilon^{(f)}/|\epsilon_0|$, for example.
The result of a corresponding study is summarized in Fig.\ \ref{fig1}.
In this diagram various regions in the $(\Delta p^2/2|\epsilon_0|,
\epsilon^{(f)}/|\epsilon_0|)$-plane are identified that
give rise to different types of behavior of the ionization spectra.

Region (I) pertains to parameters with $0<\epsilon^{(f)}<
\epsilon_{crit}^{(f)}\equiv \Delta
p^2/2+\epsilon_0$. According to Eq.\ (\ref{energy}), in this case 
classical trajectories start with negative (positive) radial momenta
$-(+)\nabla_{{\bf{x}}_0} S_0({\bf{x}}_0)$
from the positive (negative) $z$-half plane. In order to illustrate
the typical behavior of the resulting spectra we
consider as an example the parameter values $\Delta
p^2/2|\epsilon_0|=6.25$, $\epsilon^{(f)}/|\epsilon_0|=1.84$.
Figure \ref{fig2}(a) depicts the initial positions of the
classical trajectories in the $(x,z)$-plane and their respective final
angles for the initial principal quantum number $n_0=50$ (which implies
$\epsilon^{(f)}=10$ meV). To show the distribution of the scattering
angles clearly, in Fig.\ \ref{fig2}(b) the ``deflection function" is
displayed, i.e., the dependence of the final angle $\Theta_{f}$ on the
polar angle $\Theta_{i}$ of the corresponding initial position.
From Figs.\ \ref{fig2}(a) and (b) it is apparent that three
different initial positions pertain to each final emission angle $\Theta_{f}$.
The relative significance of these trajectory classes
for the ionization signal can be inferred from Fig.\ \ref{fig2}(c) where
their classical weights
$\tilde{P}_{cl}^2=| P_{cl}|^2/|A_{cl}({\bf x}_0)|^2$
(i.e., the Jacobi determinants) are depicted as a function of the
final angle $\Theta_f$. For small and very large final angles the
contributions of all three classes are relevant whereas in the
intermediate region the weight of class (II) becomes negligibly
small. This decrease is due to the fact that class (II) includes
trajectories with initial positions ${\bf x}_0$ in the vicinity of
$\bf x=\bf 0$ and the behavior of such trajectories depends very sensitively
on the initial conditions. As shown in Fig.\ \ref{fig2}(d)
the existence of different contributing trajectories with approximately
equal weight leads to pronounced interference effects in the ionization
spectrum. In particular, for small angles one observes rapid oscillations
which gradually diminish as $\Theta_f$ is increased. This diminution is
directly related to the peculiar $\Theta_f$-dependence of the classical
weight of trajectory class (II) that was described above.

An important feature of Figs.\ \ref{fig2}(a) and (b) is the existence of
trajectories with $\Theta_f = 0^{\circ}
(180^{\circ})$ whose initial positions do not lie on the $z$ axis.
Consequently, due to the rotational symmetry of the system around this axis,
there are infinitely many classical trajectories that contribute
to the ionization amplitude at these final angles so that the classical
probabilities $|P_{cl}|^2$ diverge there. This phenomenon
is known as semiclassical glory scattering \cite{Berry}.
For a proper semiclassical description the primitive ionization amplitude of
Eq.~(\ref{semi}) has to be regularized with the help of uniformization
methods \cite{Berry}.
Thereby, one replaces the contributions of the trajectory classes
$a$ and $b$ at both sides of the divergence by the uniform expression
\begin{eqnarray}
&&e^{i[(f_a + f_b)/2 - {\tilde \mu}_b \pi/2 - \pi/4]}
\sqrt{ \frac{\pi(f_a - f_b)}{4}}
\{ (g_a + g_b)\times\nonumber\\
&& J_0[(f_a - f_b)/2] + i (g_a - g_b) J_1[(f_a - f_b)/2] \}
\label{uni1}
\end{eqnarray}
with $f_j = S_{j} + W_j({\bf x}_0^{(j)})$ and $g_j = P_j^{(cl)}$ ($j=a,b$).
The trajectories are labeled in such a way that $f_a > f_b$ so that
${\tilde \mu}_a={\tilde \mu}_b +1$.
The Bessel functions of order zero and one are
denoted $J_0(x)$ and $J_1(x)$.
The contribution of the third trajectory can be described by the
primitive semiclassical expression given in Eq.~(\ref{semi}).
In Fig.\ \ref{fig2}(d) the quantum mechanical ionization spectrum (full
curve) is compared to the uniform (dashed) and the primitive (dotted)
semiclassical approximation. All spectra in this section are calculated
for an initial hydrogenic energy eigenstate with $l_0=0$.
The uniform curve describes the quantum
mechanical result very well over the whole angular range whereas the
primitive approximation diverges in small regions around $\Theta_f =
0^{\circ}$ and $180^{\circ}$.
The characteristic qualitative features of the example described above,
i.e., the existence of three contributing trajectories for all angles
$0^{\circ}<\Theta_f<180^{\circ}$, their relative weights and the
appearance of forward and backward glory effects are characteristic for
all spectra represented by region (I) of Fig.\ \ref{fig1}.

The second region of particular importance in Fig.\ \ref{fig1} is
denoted (II). This region is bounded from below and above by the curves
\begin{equation} \label{curve1}
E^2+2(1-3\Delta)E+8\Delta\sqrt{E\Delta}-3\Delta^2-2\Delta+1=0
\end{equation}
and
\begin{equation} \label{curve2}
E^2+2(1-3\Delta)E-8\Delta\sqrt{E\Delta}-3\Delta^2-2\Delta+1=0
\end{equation}
with $E=\epsilon^{(f)}/|\epsilon_0|$ and $\Delta=\Delta p^2/2
|\epsilon_0|$ \cite{curves}. For $\Delta\geq 1$, relations (\ref{curve1}) and
(\ref{curve2}) can be approximated as
$E=\Delta-1/2$ and $E=9\Delta-3/2$, respectively. The spectra in
region (II) are characterized by the appearance of a semiclassical rainbow
scattering phenomenon and a backward glory whereas the forward glory
present in the spectra of region (I) has disappeared. Again this
behavior is most clearly illustrated by studying a particular example.
To this end, the parameter values $\Delta p^2/2|\epsilon_0|=6.25$,
$\epsilon^{(f)}/|\epsilon_0|=7.35$ are used (with $n_0=50$ in Fig.\
\ref{fig3} which implies $\epsilon^{(f)}=40$ meV). In contrast to
the previous case (and, more generally, as soon as $\epsilon^{(f)}>
\epsilon_0+\Delta p^2/2$) trajectories originating from the positive
(negative) $z$-half plane now start with positive (negative) radial momentum.
Figure \ref{fig3}(a) depicts the location of initial positions in the
$(x,z)$-plane and their corresponding final angles whereas Fig.\
\ref{fig3}(b) shows again the deflection function.
Apparently, for final angles larger than the ``rainbow angle"
$\Theta_r$ the contributions of three different trajectory classes have
to be taken into account. At $\Theta_r$, however, two of these classes
coalesce which gives rise to the semiclassical rainbow scattering
phenomenon \cite{Berry}. Again the classical weights of the
corresponding classical trajectories diverge at this angle
[cf.\ Fig.\ \ref{fig3}(c)].
To determine the transition
amplitude the primitive approximation has
to be replaced by a uniformized expression.
Using the notation of expression (\ref{uni1}) and defining
$\xi = [3(f_1 - f_2)/4]^{2/3}$
this uniform transition amplitude takes the form
\begin{eqnarray}
&&\sqrt{\pi} e^{i[(f_1 + f_2)/2 - {\tilde \mu}_2 \pi/2 - \pi/4]}
\{ (g_1 + g_2)\times\nonumber\\
&& \xi^{1/4}{\rm Ai}(-\xi) + i (g_1 - g_2)
\xi^{-1/4}{\rm Ai}'(-\xi) \}.
\label{uni2}
\end{eqnarray}
In (\ref{uni2}) ${\rm Ai}(\xi)$ and ${\rm Ai}'(\xi)$ denote the Airy
function and its derivative. Quantum mechanically, the rainbow
scattering phenomenon does not cease abruptly at the rainbow angle but
extends smoothly into the classically forbidden region. This may be taken
into account semiclassically by extrapolating the quantities $f_i$,
$g_i$, and $\xi$ of expression (\ref{uni2}) to values $\Theta_f<\Theta_r$.

Furthermore, Figs.\ \ref{fig3}(a)-(c)
show that the forward glory has disappeared but that the
backward glory is still present.
The classical weight associated with the different
types of trajectories can be inferred from Fig.\ \ref{fig3}(c). Figure
\ref{fig3}(d) shows that the most prominent difference between the spectra
of cases (I) and (II) is that the latter start for
small angles with a smooth non-oscillating part. Oscillations set in
only around $\Theta_r$.

If, for a fixed value of $\Delta p^2/2|\epsilon_0|$, the final energy
$\epsilon^{(f)}/|\epsilon_0|$ is increased the rainbow angle $\Theta_r$
grows towards the value of $180^{\circ}$ and the interference effects in
the spectrum are more and more diminished. If the line given
by Eq.\ (\ref{curve2}) is crossed, rainbow and backward glory scattering
vanish completely. In this region (IV) the spectrum is determined by
the contributions of a single trajectory class and interferences are no
longer visible. However, Fig. \ref{fig1} also indicates
that in this dynamical region
the angle-integrated ionization probability $dP_{ion}/d\epsilon^{(f)}$ is
very small so that this type of behavior is only of little practical relevance,
in general.

Regions (I) and (II) are separated by a small area (III) in which
the classical dynamics is more complicated. In this
transition region besides the backward glory both forward glory and
rainbow scattering characteristic of regions (I) and (II) are present.
In addition, however, a second classical rainbow angle appears at a very
small value of $\Theta_f$. A proper semiclassical description of this
behavior would require uniformization based on a higher-order
catastrophe \cite{Berry}.

Figures \ref{fig2}(a) and \ref{fig3}(a) indicate that there is a strong
correlation between the final direction into which the electron is
emitted and its position 
before application of the pulse. This correlation might be useful for
probing the spatial distribution of localized wave packets. In order to
assess the perspectives of such an approach we plot in Fig.\ \ref{fig6}
several ionization spectra obtained in the sudden-impact approximation.
They pertain to a radial Rydberg wave packet detected at different stages
of its time
evolution. For the calculation, the wave packet was assumed to be
generated by a weak and short laser pulse as described in Ref.\
\cite{ARZ86} and the electronic angular momentum was taken as $l_0=0$
to show the effects in question most clearly. The spectral envelope
$\tilde{\cal E}(\epsilon)$ 
of the laser pulse was chosen as $\tilde{\cal E}(\epsilon)\propto
\exp[-(\epsilon-{\bar \epsilon})^2(\delta t)^2]$
with the mean excited energy ${\bar \epsilon}=-2\times 10^{-4}$ a.u.\  
corresponding to a
mean quantum number ${\bar n}=50$ and an energy spread determined by
$\delta t=0.035\ T_{cl}$. Figure \ref{fig6}(a) shows the spectra (on a
non-logarithmic scale) for $\epsilon^{(f)}=40$ meV and various delay
times between the exciting laser pulse and the HCP with $\Delta
p=0.05$ a.u. All delay times are chosen such that the wave packet has not
yet reached its outer turning point and is still moving outward. The
inset depicts some corresponding radial distributions $|\psi(t)|^2$ for the
wave packet. The connection between these distributions and the form of the
spectra can be made with the help of the right part of Fig.\ \ref{fig3}(a)
where $z>0$. As long as $|\psi(t)|^2$ is peaked at a distance $r_c$ less
than ca.\ 4000 a.u.\ the spectrum attains its maximum at an angle $\Theta_f>0$
whose value agrees with what can be inferred directly from $r_c$ and Fig.\
\ref{fig3}(a). For larger values of $r_c$ the maximum is reached at
$\Theta_f=0^{\circ}$ but decreases as the wave packet moves beyond 4000 a.u.

For the case of an inward moving wave packet which is examined in Fig.\
\ref{fig6} the spectra are essentially peaked around the rainbow angle
$\Theta_r$ independent of the delay time. This observation agrees again
with the classical picture obtained from Fig.\ \ref{fig3} as the
associated weight of the electron trajectories diverges at this angle.
In order to achieve a connection between the maxima in the spectrum and
the electronic probability distribution similar to Fig.\ \ref{fig6}(a)
one should investigate spectra for final energies $\epsilon^{(f)}<
\epsilon_{crit}^{(f)}$ [cf. Fig.\ \ref{fig2}(a)]. In summary, Figs.\ \ref{fig6}
demonstrate that the spatial distribution of the wave packet is indeed
reflected in the ionization spectrum. In order to obtain a complete
picture of the distribution one will have to analyze a set of spectra
with different parameters. A particular advantage of the use of HCPs,
compared, e.g., to optical methods, is that they can interact with the
wave packet at every point of its orbit.

\subsection{Finite-duration HCPs}

In this subsection it is discussed to which extent the results of Sec.\
III A can be applied to the study of angle- and energy-resolved ionization
spectra from HCPs of finite duration. It is natural to examine this problem
within a semiclassical approach as the theoretical
framework discussed in Sec.\ II can be applied
equally well to both instantaneous and time-dependent pulses and its
quantitative accuracy has been demonstrated in Sec.\ III A. Thus the main task 
consists in studying the classical electronic dynamics. In
a fully quantum-mechanical treatment the partial wave expansion of Sec.\
II B would no longer be applicable. One would have to turn to a
numerical integration of the time-dependent Schr\"odinger equation which
is significantly more complicated due to the Coulomb singularity
at the nucleus and due to the large spatial extension of highly excited
Rydberg states.

In the present context our main question is to which extent the essential
results of the sudden-impact approximation can be recovered in experiments
with finite-duration HCPs. It is therefore reasonable to concentrate the
study on short pulses with durations $\tau$ up to the order of, say, one tenth
of the Rydberg electron's orbit time $T_{cl}$. Typically this corresponds to
pulse lengths around 1 ps which can easily be produced in the laboratory.
Before discussing in detail an example that illustrates characteristic effects
originating from finite-duration HCPs let us briefly summarize the
main results of our investigations:

(i) The behavior of trajectories starting close to the nucleus is modified
significantly even by very short pulses whereas the dynamics of orbits
originating at a large distance from the core is changed only very
gradually. This is of course due to the different initial velocities of
the trajectories: far way from the nucleus the electron moves slowly and
the pulse can still be considered as almost instantaneous. From Figs.\
\ref{fig2} and \ref{fig3} it follows that the part of the spectrum
with small final angles $\Theta_f$, where the ionization probability is
concentrated and where forward glory and rainbow scattering can be
observed, is related to trajectories with large initial distances
from the nucleus. Thus these essential regions of the spectrum can be
expected to be relatively insensitive to effects of finite pulse
durations. On the other hand, the behavior of the spectrum at large
final angles is determined by trajectories starting close to the core.
In this region of the spectrum, where the ionization probability is
relatively small, pulse duration effects are most prominent. In
particular, in our examples we even find backward glory scattering to vanish.

(ii) The deviations of the spectrum from the sudden-impact 
approximation are mainly determined by the pulse duration whereas 
details of the pulse shape seem to be of less importance. We
compared the classical dynamics and spectra resulting from two types of HCPs
the shapes of which were almost rectangular and a half-period sine wave,
respectively. These choices should represent typical variations of
realistic HCP shapes.
The differences in the resulting spectra for these two pulse forms
(calculated for identical pulse durations and
integrated field strengths)
were small in comparison to the deviations from the sudden-impact
approximation (see Figs.\ \ref{fig4}(c) and \ref{fig5}).

(iii) In the sudden-impact approximation there are classical
trajectories for {\it all} values of final energies
$0 < \epsilon^{(f)}<\infty$.
With increasing pulse duration, however, we find
that the range of accessible final energies becomes more and more
reduced in accordance with the energy-time uncertainty principle
mentioned in the introduction.

In order to discuss effects of finite pulse duration in more detail
we consider as an example the ionization spectrum for an HCP with
an almost rectangular pulse shape given by ${\bf E}(t)={\bf e}_z E_0
{\rm exp}[-(t/t_0)^8]$. The parameters $E_0$ and $t_0$ are determined so
that the integrated field strength and the pulse duration fulfill
$\Delta p^2/2|\epsilon_0|=6.25$ and $\tau/T_{cl}=0.05$, respectively.
Again, for the calculation
an initial hydrogenic energy eigenstate with $l_0=0$ and with
principal quantum number $n_0=50$ was chosen.
Figure \ref{fig4} shows an analysis of the corresponding
classical dynamics and of the
resulting ionization spectrum for a final energy of
$\epsilon^{(f)}/|\epsilon_0|=7.35$ (compare to Fig.\ \ref{fig3}).
Figure \ref{fig4}(a) details the changes in the classical dynamics.
Three major modifications can be observed:

(i) The final angles $\Theta_f$ of trajectories with initially outgoing
radial momentum do not cover the full range $0^{\circ}<\Theta_f<90^{\circ}$
as in the sudden-impact approximation. Instead, they are restricted to
values $0<\Theta_f<\Theta_m$ with $\Theta_m\approx 23.7^{\circ}$ in the
present case. Here, this does not lead to major changes in the spectrum.
This is because in the angular region where the deviations of this trajectory
class from the sudden-impact case become significant the dominant
contributions already originate from trajectory classes (I) and (II)
[see Figs.\
\ref{fig3}(b) and \ref{fig4}(b)]. However, due to the absence of trajectory
class (III) for $\Theta_f>\Theta_m$ the small short-period modulations
of the spectrum of Fig.\ \ref{fig3}(c), which are
visible for $\Theta_f$ between $25^{\circ}$ and
$55^{\circ}$, are no longer present in Fig.\ \ref{fig4}(c).

(ii) A new class of trajectories appears
which has no counterpart in the sudden-impact approximation. These orbits
have initially incoming radial momenta
and originate from the half-plane $z>0$. Their final angles
vary between $0^{\circ}$ and $27.5^{\circ}$. The maximum final angle is
not attained at $(x=0,z=0)$ but at a finite distance from the nucleus so
that there is a rainbow effect connected with this class of trajectories
[cf. Fig.\ \ref{fig4}(b)].
However, due to the low classical weight of these trajectories they
show up in the spectrum
most prominently
in the form of small modulations in the
vicinity of $\Theta_f=0^{\circ}$.

(iii) The initial positions of trajectories which start with incoming
radial momenta from the half-plane $z<0$ are changed significantly. In
addition, the final angle $\Theta_f=180^{\circ}$ is reached only for
initial positions with $z=0$, so that backward glory scattering has
disappeared. The changes in the ionization spectrum at large angles
which are due to these effects represent the most significant influence of
finite pulse durations.

In Fig.\ \ref{fig4}(c) the resulting ionization spectrum (full curve)
is compared 
to a spectrum which has been calculated for an HCP of identical
integrated field strength and pulse duration but with shape
${\bf E}(t)={\bf e}_z E^{\prime}_0 \sin(2\pi t/\tau^{\prime}),t\leq
\tau^{\prime}/2$ (dashed curve).
The spectrum for the sudden-impact approximation is also shown (dotted curve).
As mentioned above, for these pulse durations the influence of the
HCP shape is only secondary even for large final angles. However, as
the pulse length is increased to values larger than ca.\ 0.1 $T_{orb}$
these influences eventually become more significant. Figure \ref{fig5} shows
the results of an investigation similar to Fig.\ \ref{fig4}(c) for a final
energy of $\epsilon^{(f)}/|\epsilon_0|=3.68$ ($\epsilon^{(f)}=20$ meV)
which corresponds to
region (I) of the sudden-impact approximation (cf.\ Fig.\ \ref{fig1}). As in
the previous case, it is found that the main features of the spectrum of
the sudden-impact approximation are left almost unchanged. The
classical dynamics change similarly to the previous example.
In particular, the new trajectory class still exists, but
in the rapidly oscillating part of the spectrum
its contribution is almost undiscernible.

\section{Summary and conclusions}

A general semiclassical treatment of excitation of weakly bound
Rydberg electrons by intense HCPs has been presented. Thus
a quantitative connection between observable transition probabilities and
the underlying classical dynamics of the excited Rydberg electron
has been established. This approach is numerically
accurate for high principal quantum numbers of the
Rydberg electron and in cases in which the energy transfer from the exciting
HCP to the Rydberg electron is small in comparison with typical
ionization energies of low-lying states.
The underlying classical dynamics of the Rydberg electron yields a clear 
and natural explanation for the oscillatory structures which govern the
resulting energy- and angle-resolved ionization spectra. These structures
arise from interference between probability amplitudes associated with
different ionizing
trajectories. Thereby two types of semiclassical catastrophes appear, namely
glory scattering  for electron emission in the forward and backward direction
of the polarization of the HCP and a rainbow phenomenon.
Parts of these oscillatory structures depend strongly on the ratio between the
pulse duration of the exciting HCP and the classical Kepler period of the
initially prepared Rydberg state. The amplitude and phase information which
is contained in the semiclassical probability amplitudes
might be useful for the reconstruction of quantum states of Rydberg electrons.
A first indication of the usefulness of such an approach was given by
the investigation of the spectra pertaining to radial wave packets.

G.~A.\ acknowledges support by the Deutsche For\-schungs\-ge\-mein\-schaft
within the SPP ``Zeit\-ab\-h\"an\-gige Ph\"anomene und Methoden." O.~Z.\ is
supported by the U.S.\ Office of Naval Research under Contract No.\ 14-91-J1205
and by the U.S.\ Army Research Office.

\newpage
\centerline{\psfig{figure=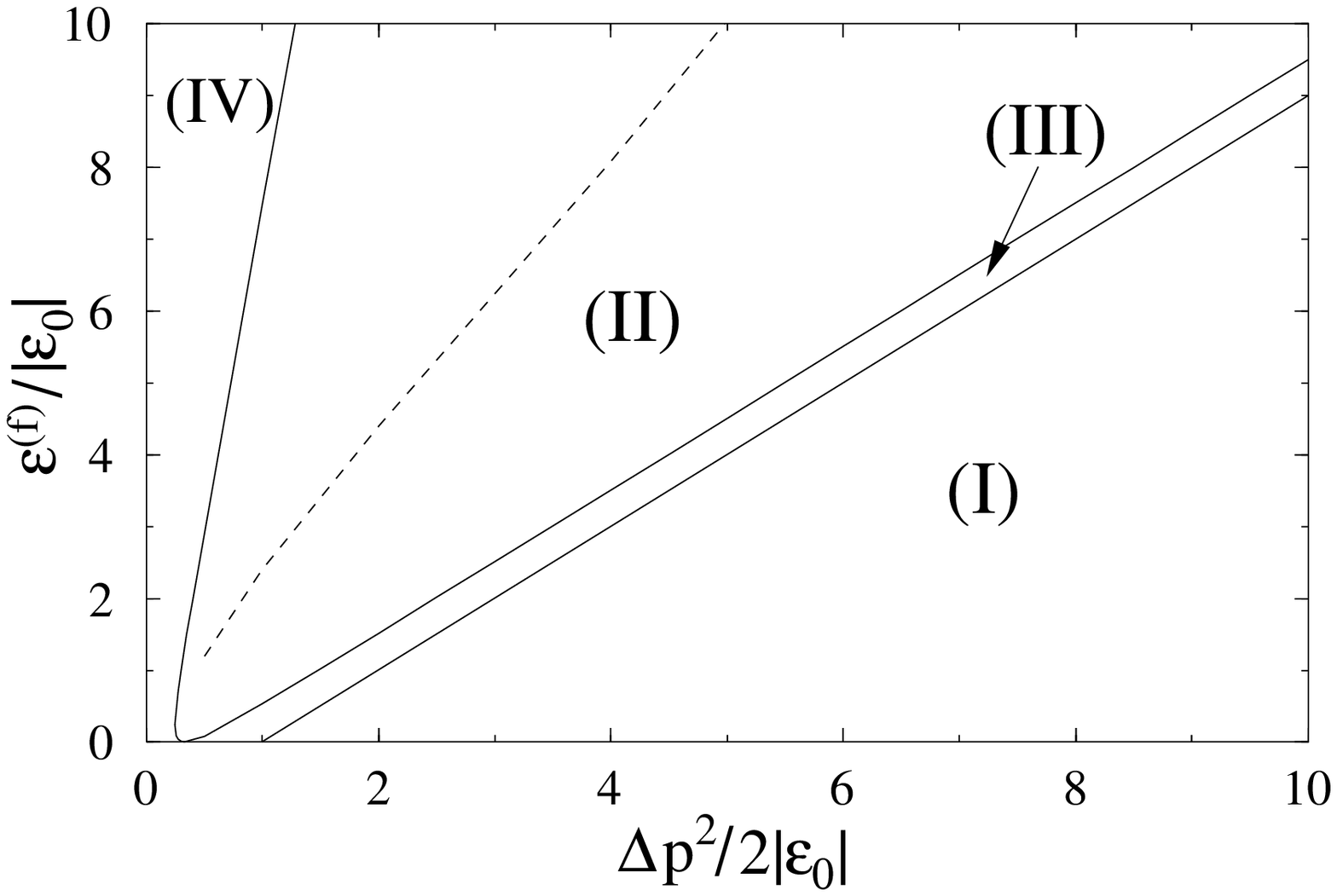,width=8.6cm,clip=}}
\begin{figure}
\caption{Classification of ionization spectra in the sudden-impact
approximation. The following semiclassical catastrophes are observed
in the respective areas of the plane spanned by $\Delta p^2/2|\epsilon_0|$
and $\epsilon^{(f)}/|\epsilon_0|$:
(I) forward and backward glory, (II) rainbow
and backward glory, (III) two rainbows, forward and backward glory, (IV)
no semiclassical catastrophes. For a given value of
$\Delta p^2/2|\epsilon_0|$
the ionization probability $dP_{ion}/d{\epsilon^{(f)}}$ is concentrated
below the dashed line [in the sense that $dP_{ion}/d{\epsilon^{(f)}}
<0.025 (dP_{ion}/d\epsilon^{(f)})_{max}$, approximately, beyond this
line].}
\label{fig1}
\end{figure}

\centerline{\psfig{figure=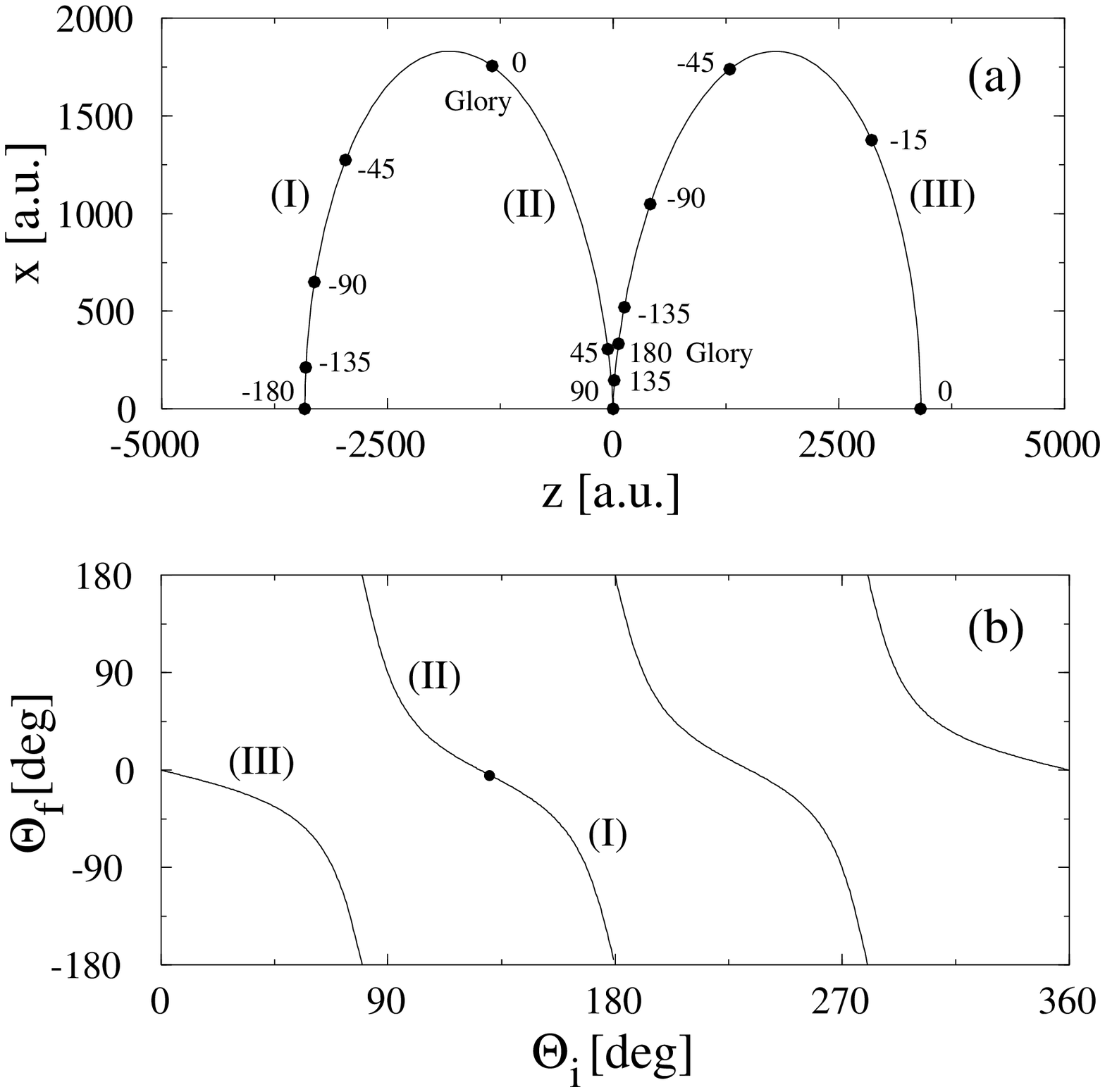,width=8.6cm,clip=}}
\centerline{\psfig{figure=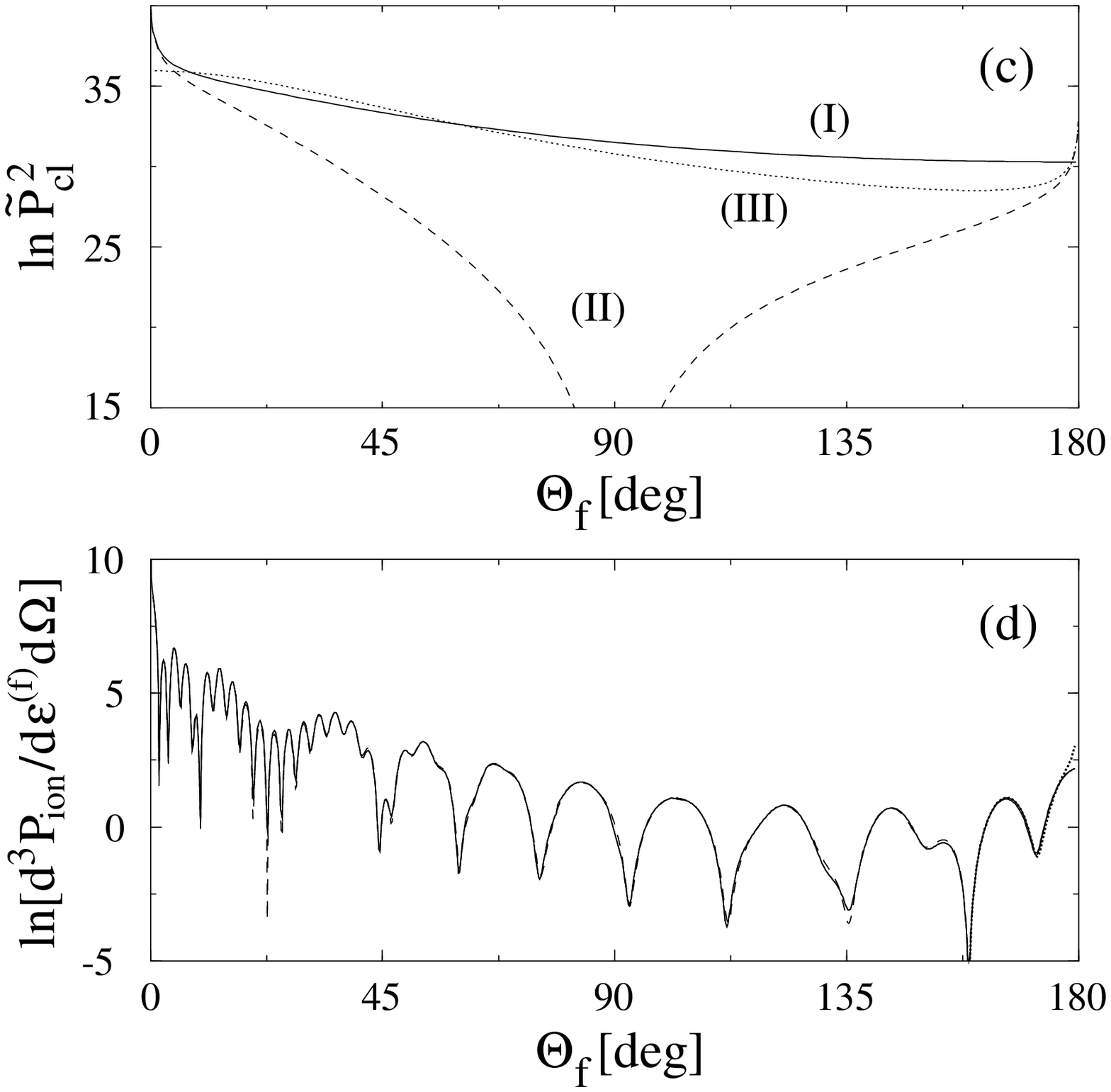,width=8.6cm,clip=}}
\begin{figure}
\caption{
(a) Initial positions of classical trajectories with $\Delta p^2/
2|\epsilon_0|=6.25$, $\epsilon^{(f)}/|\epsilon_0|=1.84$, $n_0=50$
according to Eq.\ (\protect\ref{energy}). The numbers indicate final
emission angles $\Theta_f$ for specific initial positions.
Angles are counted
counterclockwise from the $z$ axis. The figure has to be continued
into three dimensions by rotation around the $z$ axis.
The roman numbers relate the different trajectory classes to diagram (c).
Axes are labelled in atomic units.
(b) Final angle $\Theta_f$ as a function of the polar angle $\Theta_i$
of the corresponding initial position.
(c) Classical weight
$\tilde{P}_{cl}^2=| P_{cl}|^2/|A_{cl}({\bf x}_0)|^2$ (in a.u.) as a
function of $\Theta_f$ for the trajectory classes distinguished in (a).
(d) Angular distribution of the ionized electron
$\ln\{d^3P_{ion}/d\epsilon^{(f)} d\Omega [{\rm a.u.}] \}$ in the
sudden-impact approximation:
quantum mechanical (full), uniform semiclassical (dashed)
and primitive semiclassical (dotted) result.}
\label{fig2}
\end{figure}

\centerline{\psfig{figure=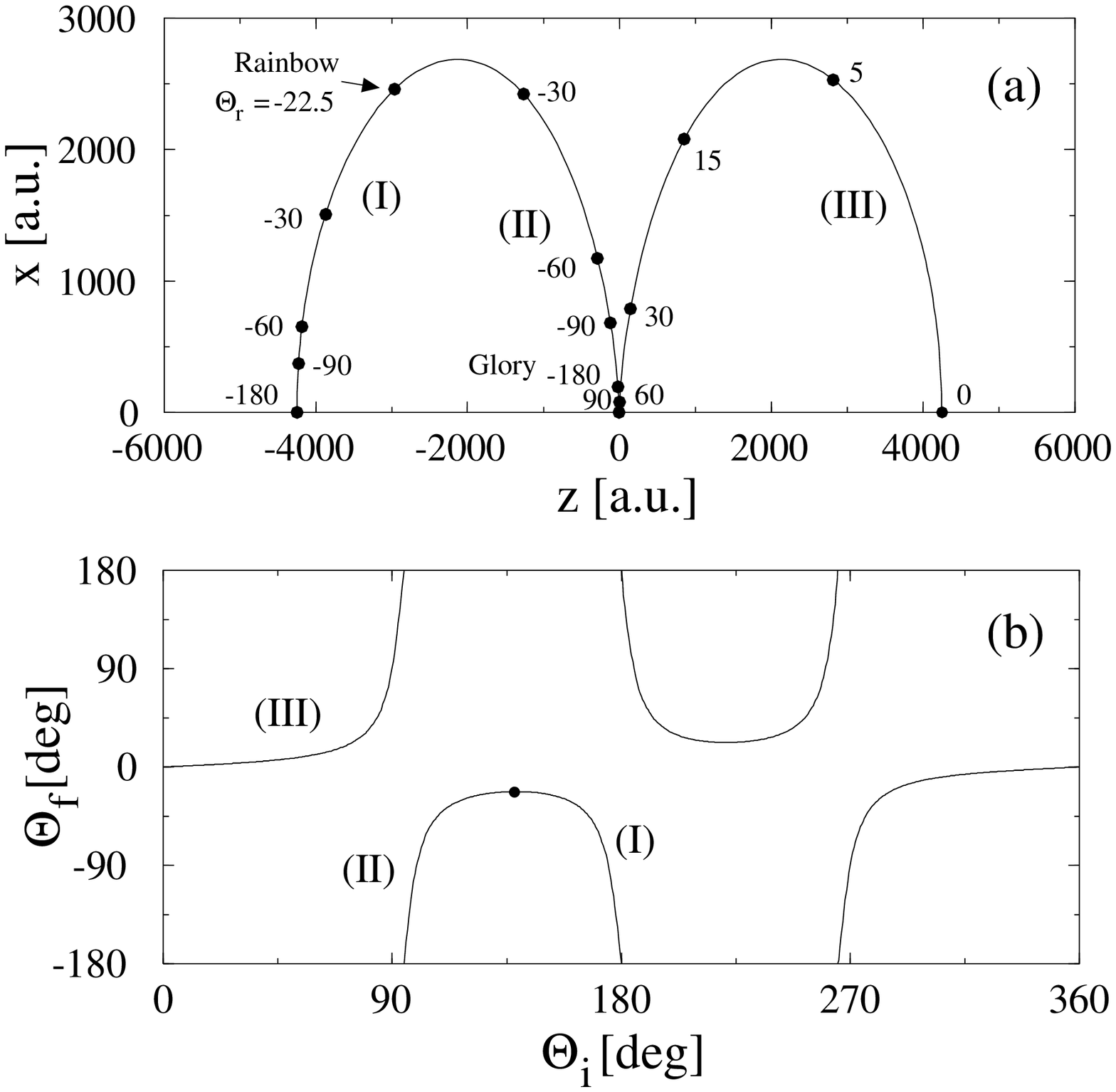,width=8.6cm,clip=}}
\centerline{\psfig{figure=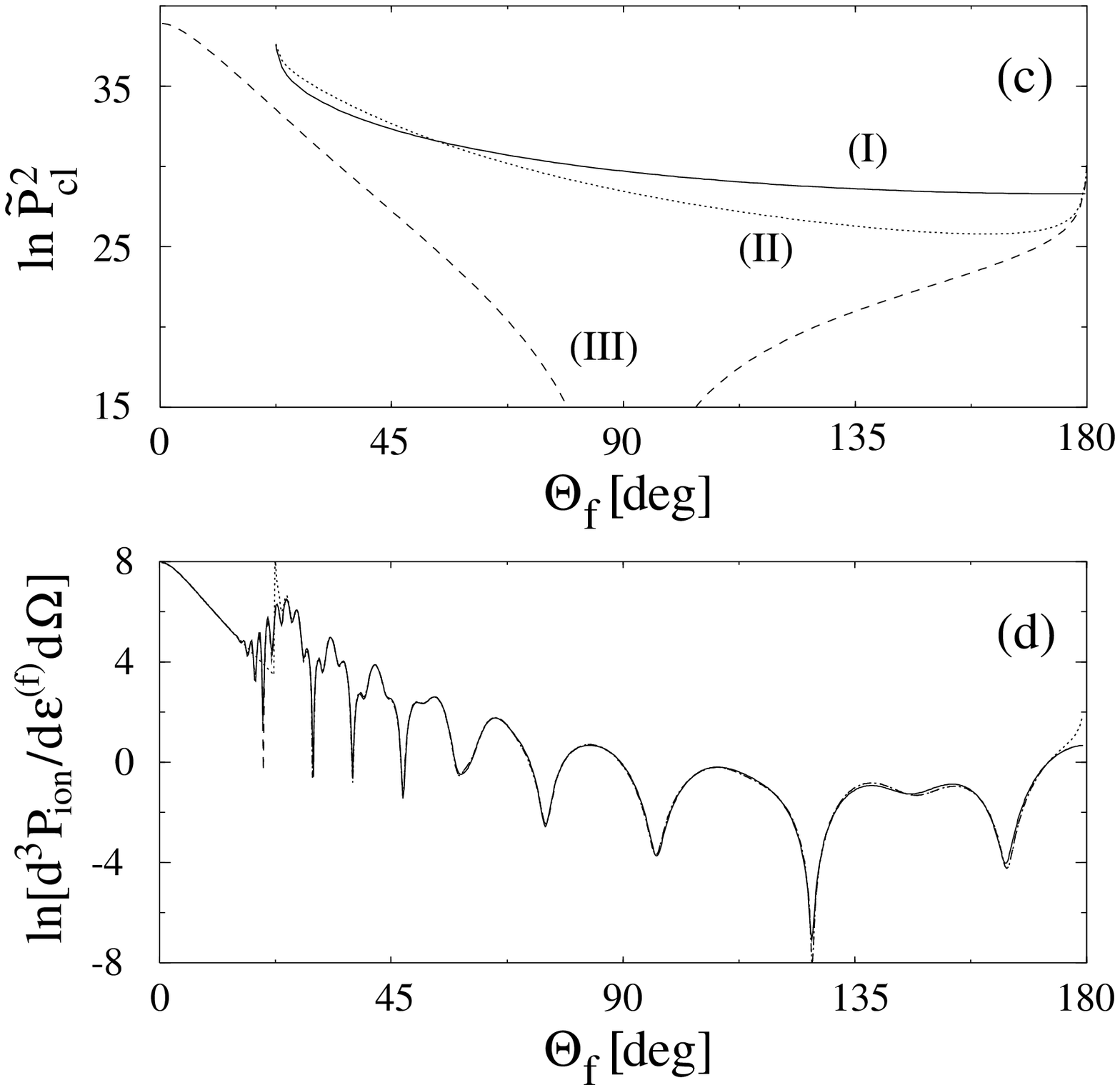,width=8.6cm,clip=}}
\begin{figure}
\caption{
Same as Fig.\ \protect\ref{fig2} for $\epsilon^{(f)}/|\epsilon_0|=7.35$.}
\label{fig3}
\end{figure}

\centerline{\psfig{figure=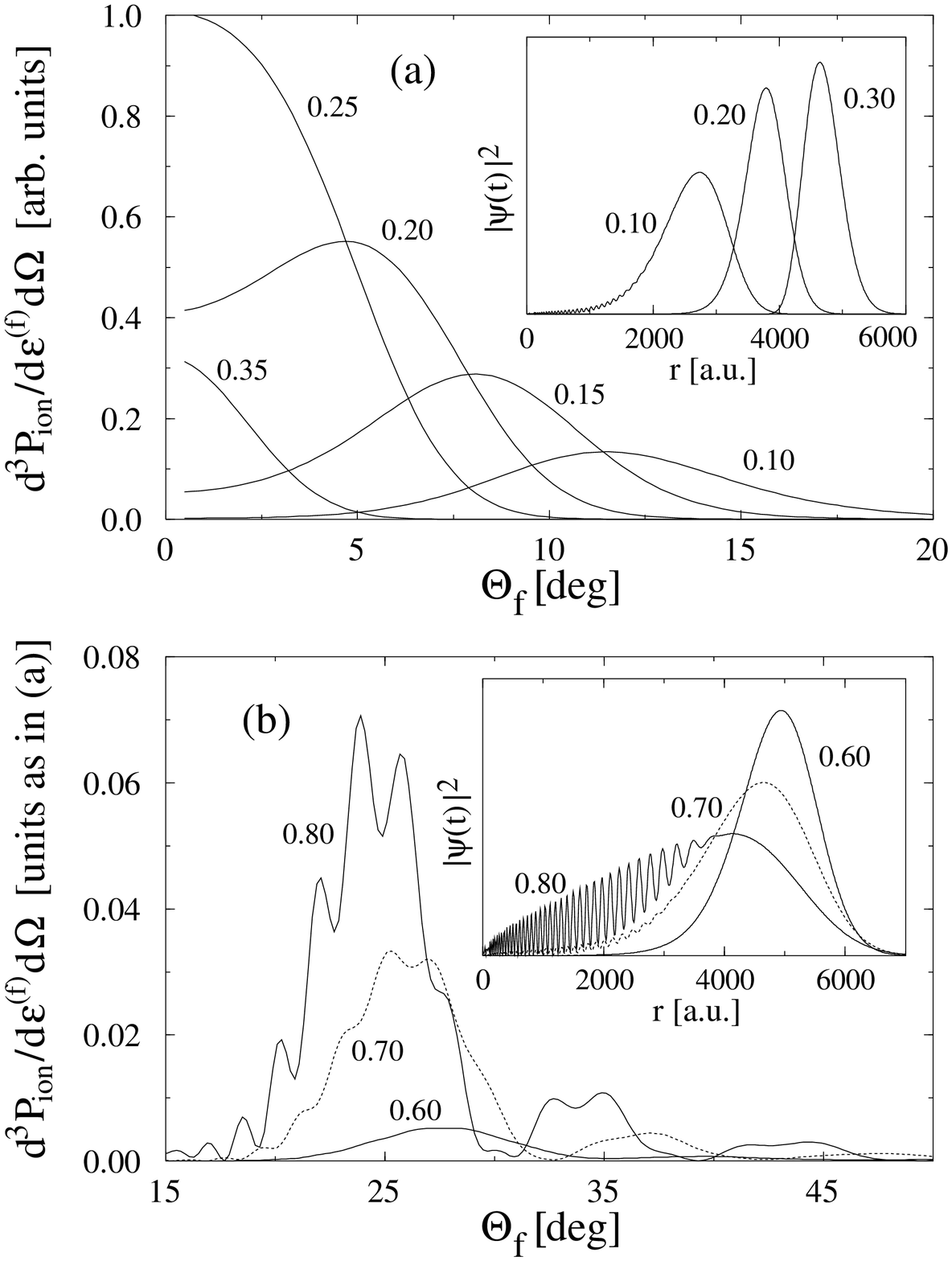,width=8.6cm,clip=}}
\begin{figure}
\caption{
Angle- and energy-resolved ionization spectra obtained from radial Rydberg
wave packets in the sudden-impact approximation. (a) Outward travelling
wave packets; the respective delay times between excitation and probing
are indicated in units of the classical wave packet orbit time $T_{cl}=20$ ps.
(b) Incoming wave packets. Radial distributions of the wave packet at
different times are shown in the insets. All relevant parameter values are
given in the text.}
\label{fig6}
\end{figure}

\centerline{\psfig{figure=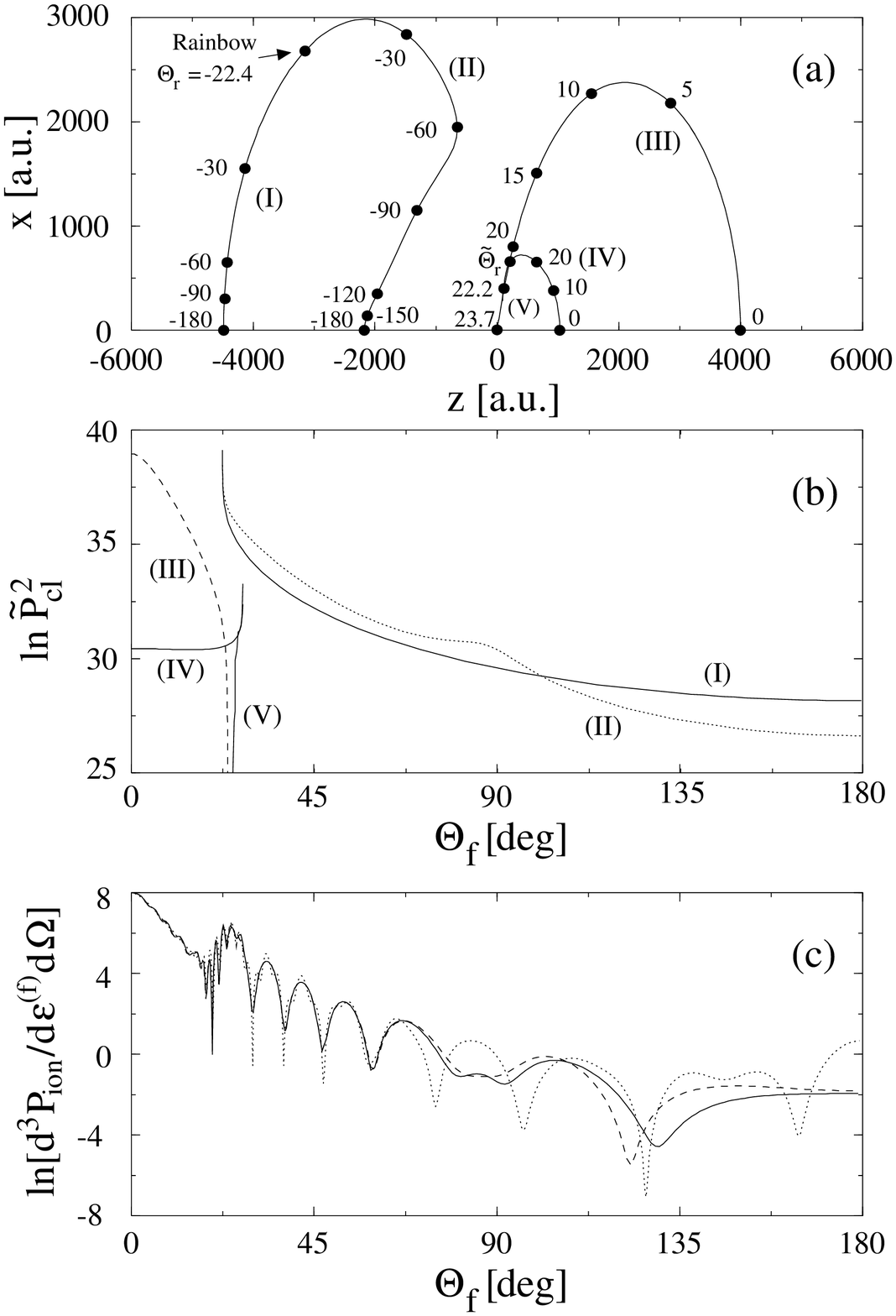,width=8.6cm,clip=}}
\begin{figure}
\caption{
Same as Fig.\ \protect\ref{fig3} but now for the
quasi-rectangular HCP of duration $T_p=0.05 T_{orb}$ described
in the text (deflection function not shown).
(a) The new trajectory classes which are denoted (IV) and
(V) are separated by the ``rainbow point" $\tilde \Theta_r$. The initial
points for (V) almost coincide with those for (III) (but the
initial radial momenta are opposite). Indicated angles refer to (III).
(c) Full curve: ionization probability for the pulse described above.
Dashed: sine-shaped pulse as described in the text. Dotted: result of
the sudden-impact approximation.}
\label{fig4}
\end{figure}

\centerline{\psfig{figure=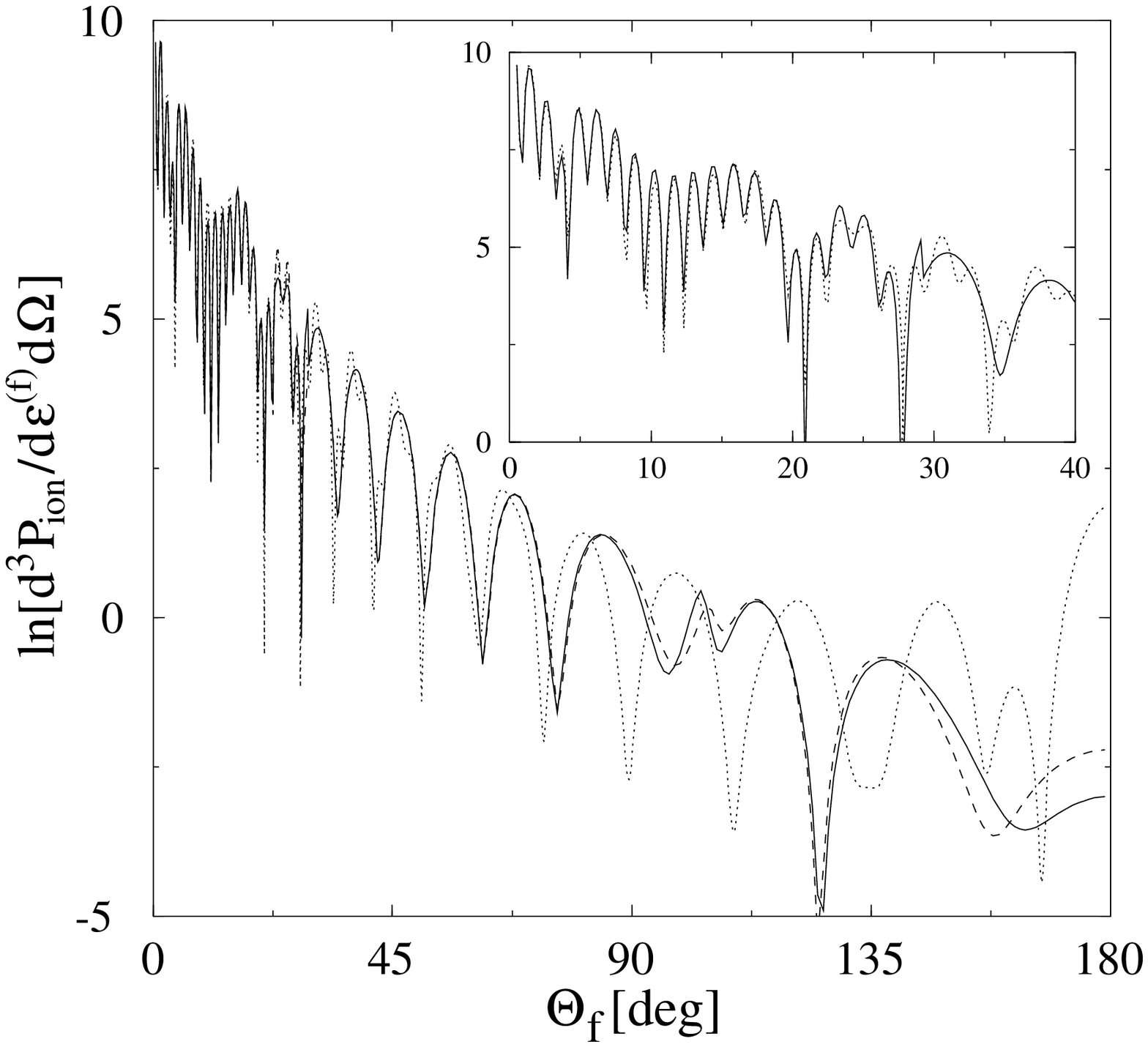,width=8.6cm,clip=}}
\begin{figure}
\caption{
Same as Fig.\ \protect\ref{fig4}(c) for $\epsilon^{(f)}/|\epsilon_0|=3.68$.
The inset shows the spectra for the quasi-rectangular pulse described
in the text (full curve) and for the
sudden-impact approximation (dotted curve) at small angles.}
\label{fig5}
\end{figure}

\begin{references}
\bibitem{curves} Relations (\ref{curve1}) and (\ref{curve2}) can be
derived by noting that $\partial \Theta_f/\partial x_0(z_0=0)|_{
\epsilon^{(f)}=const.}$ changes sign when the boundary between regions
(II) and (III) [or (IV), respectively] is crossed.
\end{references}
\end{document}